\documentclass[prb, twocolumn, preprintnumbers,amsmath,amssymb]{revtex4-1}
\usepackage{graphicx}
\usepackage{dcolumn}
\usepackage{bm}
\usepackage{subfigure}
\usepackage{multirow}
\usepackage{amsmath}
\usepackage{color,soul}
\usepackage[font=footnotesize, justification=raggedright, singlelinecheck=false]{caption}
\usepackage[dvipsnames]{xcolor}
\usepackage{mathrsfs}
\usepackage{mathtools}
\usepackage{relsize}

\usepackage[all,cmtip]{xy}
\usepackage[normalem]{ulem}

\begin{document}

\title{ Transformable Super-Isostatic Crystals Self-Assembled from Segment Colloidal Rods}

\author{Ji-Dong Hu$^{1}$, Ting Wang$^{2}$, Qun-Li Lei$^{1, \dag}$,  and Yu-qiang Ma$^{1}$ }
 \affiliation{
   $^1$ National Laboratory of Solid State Microstructures and Department of Physics, Collaborative Innovation Center of Advanced Microstructures, Nanjing University, 210093 Nanjing, China\\
   $^2$    School of Chemistry and Life Sciences, Nanjing University of Posts and Telecommunications, 210023 Nanjing, China\\
  $\dag$ E-mail: lql@nju.edu.cn (Q.-L. Lei); 
  }

\begin{abstract}
{\textbf{Abstract}}: Transformable mechanical structures can switch between distinct mechanical states. Whether this kind of  structures can be self-assembled from simple building blocks at micro-scale is a question to be answered. In this work, we propose a self-assembly strategy for these structures based on a nematic monolayer of  segment colloidal rods with lateral cutting. By using  Monte Carlo simulation, we find that rods with different cutting degrees can self-assemble into different crystals characterized by bond coordination $z$ that varies from 3 to 6. Among them, we identify a transformable super-isostatic structure with \emph{pgg} symmetry and redundant bonds ($z=5$). We show that this structure can support either soft bulk modes or soft edge modes depending on its Poisson's ratio which can be tuned from positive to negative through a uniform soft deformation.  We also prove that the bulk soft modes are associated with states of self-stress along the direction of zero strain during the uniform soft deformation. The self-assembled transformable  structures may act as mechanical metamaterials with potential applications in micro-mechanical engineering.

\end{abstract}

\maketitle

Under thermodynamic equilibrium, particles in colloidal suspensions can self-assemble into ordered structures driven by either entropy or enthalpy, which is not only  an interesting  topic in physics, but also has potential {applications} in  {the} fabrication of functional materials in large scale.~\cite{manoharan2015colloidal,boles2016self, moon2010chemical} 
In colloidal suspensions,  the interaction between colloidal particles can be adjusted by modifying physical/chemical properties of colloidal surface or solvents.~\cite{curk2021charge,koshkina2021surface} Moreover, with the  development of particle synthesis technology,~\cite{lotito2022playing} a variety of non-spherical colloids with uniform size and shape can be prepared, including  rods, ellipsoids, polyhedrons, bowl-shape and deformed sphere particles.~\cite{siavashpouri2019structure,rao2020leveraging,
neophytou2021facile,henzie2012self,sacanna2013shaping,gong2017shape,
forster2011assembly,meijer2017observation,
miszta2011hierarchical,kraft2012surface,he2020colloidal,
li2020colloidal,edmond2021large,lim2023engineering}  In principle, these complex colloidal particles can self-assemble into crystals with arbitrary space symmetry or even quasi-crystals.~\cite{dotera2014mosaic,malescio2022self} Due to the large parameter space and complex many-body effects, currently, researchers have to combine the experiments and computer simulations to fully understand the self-assembly mechanism and  physical properties of the self-assembled structures.~\cite{haji2009disordered,kraft2012surface,agarwal2011mesophase,ni2012phase,marechal2010phase,
angioletti2012re,krishnamurthy2022computer,damasceno2012predictive,lin2017clathrate,deng2020self}

Aside from  controlling the  transmission  of lights,~\cite{meade2008photonic,lei2018self,rao2020leveraging,neophytou2021self,cai2021colloidal}  self-assembled colloidal crystals can also manipulate the propagation of phonons or sounds, and exhibit special mechanical properties.~\cite{thomas2006colloidal}  For example, it was found that spherical two-end-patchy colloidal particle can self-assemble into deformable kagome lattice stabilized by vibrational entropy.~\cite{chen2011directed,mao2013entropy}  This structure is at the verge of mechanical stability since the degrees of freedom in the structure are merely  balanced by the mechanical constraint, leaving only few global soft modes. Such deformable structures are called  isostatic (Maxwell) lattices, which generally satisfy  $z=2d$, with $z$ the coordination number of the structure and $d$ the spatial dimension. ~\cite{maxwell1864calculation,lubensky2015phonons,mao2018maxwell}  Recently, it has been found that half-cylinder colloidal rods can self-assemble into self-dual isostatic crystals with hidden-symmetry, which exhibit the classical analog of Kramers degeneracy and critically tilted Dirac cone in quantum systems.~\cite{fruchart2020dualities,lei2022duality,lei2021self} On the other side, theoretical studies have shown that other isostatic structures can  exhibit topological phase transitions and robust topologically protected soft edge states.~\cite{rocklin2017transformable,sun2012surface, kane2014topological,
czajkowski2022duality,mao2018maxwell,lubensky2015phonons,ma2023nonlinear}  Considering the fact that current phononic and mechanical matamaterials can only be fabricated through top-down methods, e.g, 3D printing techniques,~\cite{shahrubudin2019overview} the self-assembly of deformable crystals  provides us another option to obtain these materials at microscale with potential applications in acoustic cloaking,~\cite{zheng2014acoustic} focusing,~\cite{yang2004focusing} shock/noise reduction,~\cite{gorishnyy2005hypersonic} and nano-mechanically engineering.~\cite{singh2021design,czajkowski2022conformal,hu2023engineering}

\begin{figure*}[htbp] 
	\resizebox{190mm}{!}{\includegraphics[trim=0.5in 0.0in 0.0in 0.0in]{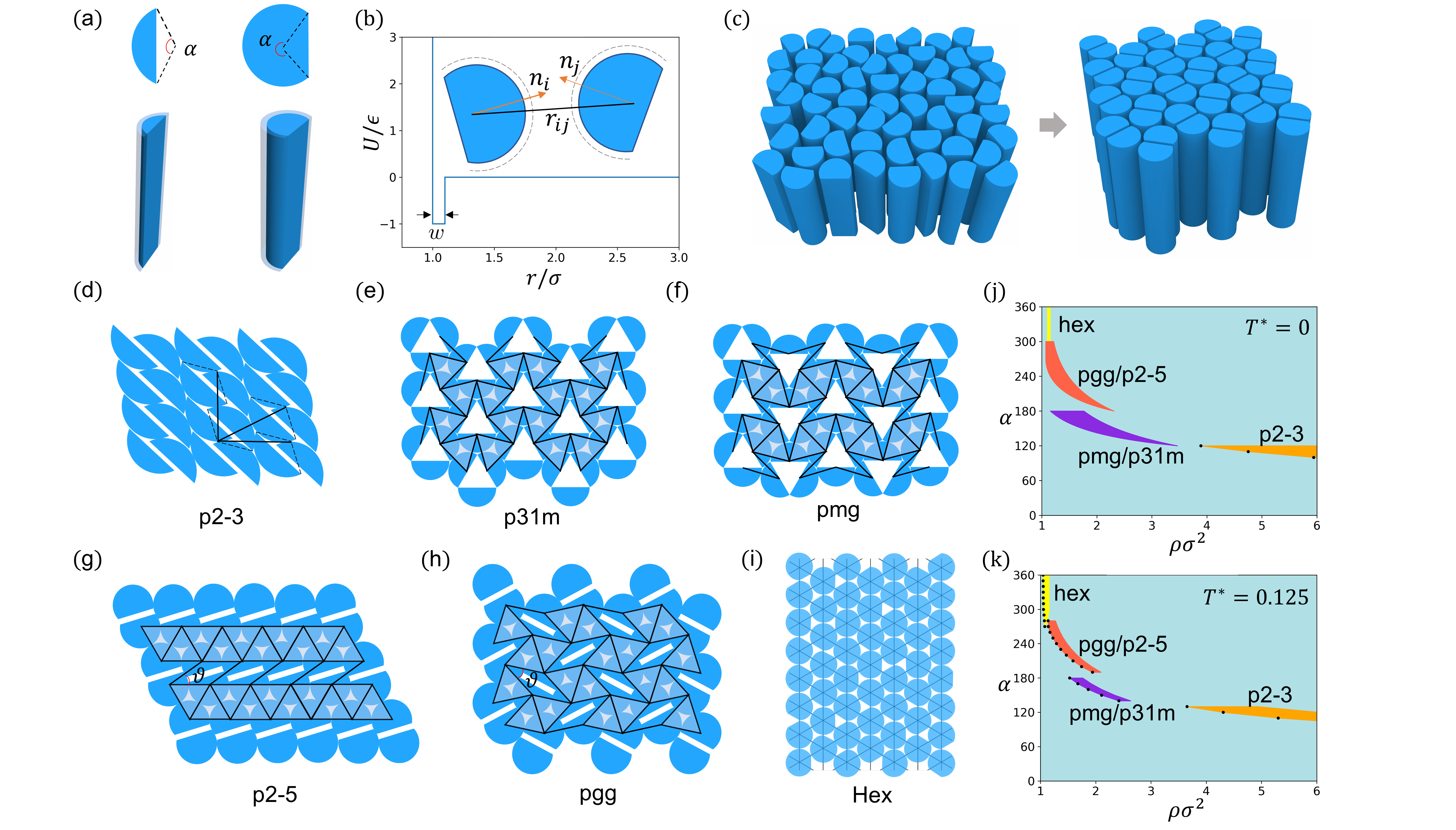} }
	\caption{  (\textbf{a}) Schematic of segment rods, where the particle shape is determined by the cutting degree $\alpha$. (\textbf{b}) Square-well potential with well width $w$ and depth $\varepsilon$ between curved surfaces from two rods. (\textbf{c}) The disordered fluid state can self-assemble into ordered crystalline states. (\textbf{d}) \emph{p2-3} crystal with coordination $z=3$. Black lines indicates the bonds. (\textbf{e})  \emph{p31m} crystal with $z=4$. (\textbf{f}) \emph{pmg} crystal with $z=4$. (\textbf{g}) \emph{p2-5} crystal with $z=5$. (\textbf{h}) \emph{pgg} crystal with $z=5$.  (\textbf{i}) Hexagonal (Hex) crystal with $z=6$.  (\textbf{j-k}) Phase diagrams in dimensions of $\alpha$ and density $\rho$ at temperature $T^*=0$ (upper) and $T^*=0.125$ (lower), respectively.}  \label{phase_diagram}
\end{figure*} 

In this work, by using Monte Carlo simulations, we systematically study the self-assembly of a nematic monolayer of perfectly aligned colloidal rods with lateral cutting. This system can be mapped into a 2D self-assembly of segment tiles. Under different cutting degree, the particles can self-assemble into different crystal phases characterized by bond coordination $z$ that varies from 3 to 6. Aside from the two previously reported isostatic crystals with $z=4$,~\cite{lei2021self} we identify {a} \emph{super-isostatic} structure with $z=5$, i.e., a \emph{pgg}-symmetry structure capable of supporting uniform soft deformation despite the existence of extensive redundant bonds. By calculating the phononic spectrum, we find that the self-assembled structure can in general support either bulk soft modes or edge soft modes depending on the Poisson's ratio which can be tuned from positive to negative through the uniform soft deformation. We also reveal that the bulk soft modes are associated with the states of self-stress (SSS) which is along the direction of zero strain during the uniform soft deformation. Our work not only proposes  a  strategy for experimentalists to self-assemble super-isostatic mechanical metamaterials, but also deepens our understanding of the floppy mechanism in super-isostatic structures.

\section*{Results}
\subsection*{Model and Simulation}
We consider a monolayer of $N$ anisotropic hard rods which can be obtained by doing lateral flat cutting to cylinder rods with diameter  $\sigma$  (Figure~\ref{phase_diagram}a). The degree of cutting, $\alpha$, controls the extent of the cutting, where $\alpha=360^\circ$ corresponds to a complete cylinder and $\alpha=180^\circ$ corresponds to a half-cylinder. We assume that all rods are perfectly aligned due to external fields or nematic interactions. ~\cite{sharma2014hierarchical,lei2018self} Therefore, this 3D system can be simplified to a 2D system with segment tiles of the same cutting degree. The 2D density is defined as  $\rho=N/A$ , where $A$ is the 2D area. We further assume there is short range attractive force  between lateral curved surfaces on the particles (Figure~\ref{phase_diagram}b). This attraction force could be depletion force,~\cite{kraft2012surface} capillary bridging force~\cite{bharti2016capillary} or Van Der Waals attraction,~\cite{ohshima2016van} which can be achieved through particle surface modification.~\cite{kraft2012surface,angioletti2016theory} For simplicity, we model the excluded volume interaction using a hard core interaction and use a square-well potential with a width $w$ and depth $\varepsilon$ to mimic the attraction force. The attractive potential between two particles $i$ and $j$ can be written as
\begin{equation}
U^{a}_{i,j} =
\begin{cases}\
-\varepsilon& \text{ if ${\mathbf n}_{i}\cdot{\mathbf r} _{ij}/\left|{\mathbf r} _{ij}\right| >\cos(\alpha/2)$ and }  \\  
~~~~~ &  \text{  ${\mathbf n}_{j}\cdot{\mathbf r} _{ji}/\left|{\mathbf r} _{ij}\right| >\cos(\alpha/2)$   and $\left|{\mathbf r} _{ij}\right|-\sigma < w $} \\
0& \text{otherwise.} 
\end{cases}
 \end{equation}
Here, $\boldsymbol{r}_{ij} = \boldsymbol{r}_{j} - \boldsymbol{r}_{i}$ with $\boldsymbol{r}_i$ or $\boldsymbol{r}_j$ the position of the axis of rod $i$ or $j$. $\boldsymbol {n}_i$ is the unit vector perpendicular to the cutting surface (see Figure~\ref{phase_diagram}b).  We  define the dimensionless temperature as $T^{*}=k_{B} T / \varepsilon$, where $k_{B}$ and $T$ are Boltzmann's constant and system temperature, respectively.  We perform Monte Carlo simulations  of the system under periodic boundary condition.~\cite{zheng2020hierarchical,lei2018self,lei2017role}  The codes of simulation is self-built based on classical Metropolis algorithms.~\cite{frenkel2001understanding}  Since rods are perfectly aligned, only 2D rotational or translational motions are allowed in the simulation. The equation of state (EOS) for the fluid and solid phases are obtained by using $10^7$ steps for equilibrium, followed by $3\times10^7$ steps sampling in NPT ensemble, with $N$ ranging from $400$ to $2000$ depending on different phases. In all our simulations, we choose $w=0.1\sigma$ if not stated otherwise. 

\subsection*{Phase Behaviors of Self-Assembly}
We conduct simulations of the system at different temperatures $T^*$ and densities $\rho$ , considering various cutting angles $\alpha$. Our findings reveal the presence of at least six ordered self-assembled phases in addition to the disordered fluid state (Figure~\ref{phase_diagram}d-i).  These ordered phases can be characterized by the average bond coordination $z$ of the particles. Each bond between two particles contributes an energy of $-\varepsilon$ under the influence of short-range attraction. We begin by examining the case of zero temperature, where these ordered phases with different bond coordinations are the ground states. We plot the corresponding zero temperature phase diagram at Figure~\ref{phase_diagram}j.  We find that when  $\alpha \lesssim 120^\circ$, crystal  with the \emph{p2} space-group symmetry ($z=3$) is the  ground state (Figure~\ref{phase_diagram}d). For $ 120^\circ   \lesssim   \alpha \lesssim 180^\circ$, the stable phases are isostatic lattices ($z=4$) with \emph{p31m} or \emph{pmg} symmetry. These two self-assembled crystals have been reported in the half-cylinder colloids systems.~\cite{lei2021self} When $ 180^\circ < \alpha < 300^\circ$, the ground states are two crystals with \emph{pgg} or \emph{p2} symmetry whose coordination are both $z=5$ with the latter  denoted as \emph{p2-5} structure. In these structures, the number of constraint is larger than the number of freedom degree. Yet, we still find that both structures are deformable structures with soft modes. Especially, \emph{pgg} structure  has a uniform soft mode, or Guest-Hutchinson (GH) mode~\cite{guest2003determinacy} similar to isostatic crystals. The \emph{p2-5} structure, on the other hand, has multiple local soft modes in one direction, thus lacking the ability of uniform deformation. For  $ \alpha > 300^\circ$, we find that hexagonal (Hex) lattice ($z=6$) with random particle orientations has the lowest energy.  In Figure~\ref{phase_diagram}k, we further construct the phase diagram of the system at finite temperature $T^*=1/8$  based on the EOS of the systems (see Figure~S1 in Supplementary Information (SI) for details). We find that at this temperature, the ordered phases are relatively destabilized and the corresponding phase regimes in the phase diagram shrink. In Figure~\ref{stability}c, we further calculate phase diagram  for the system with $253^\circ$ in the dimensions of temperature $T^*$ and density $\rho$ based on the EOS (see Figure~S2 in SI). One can find a broad temperature range in which the \emph{pgg} and \emph{p2-5}are stable phases at high density compared with the hexagonal crytal phase.

\subsection*{Stability Analysis of \emph{pgg} and \emph{p2-5} Structures}

\begin{figure*}[htbp] 
	\resizebox{180mm}{!}{\includegraphics[trim=0.0in 0.0in 0.0in 0.0in]{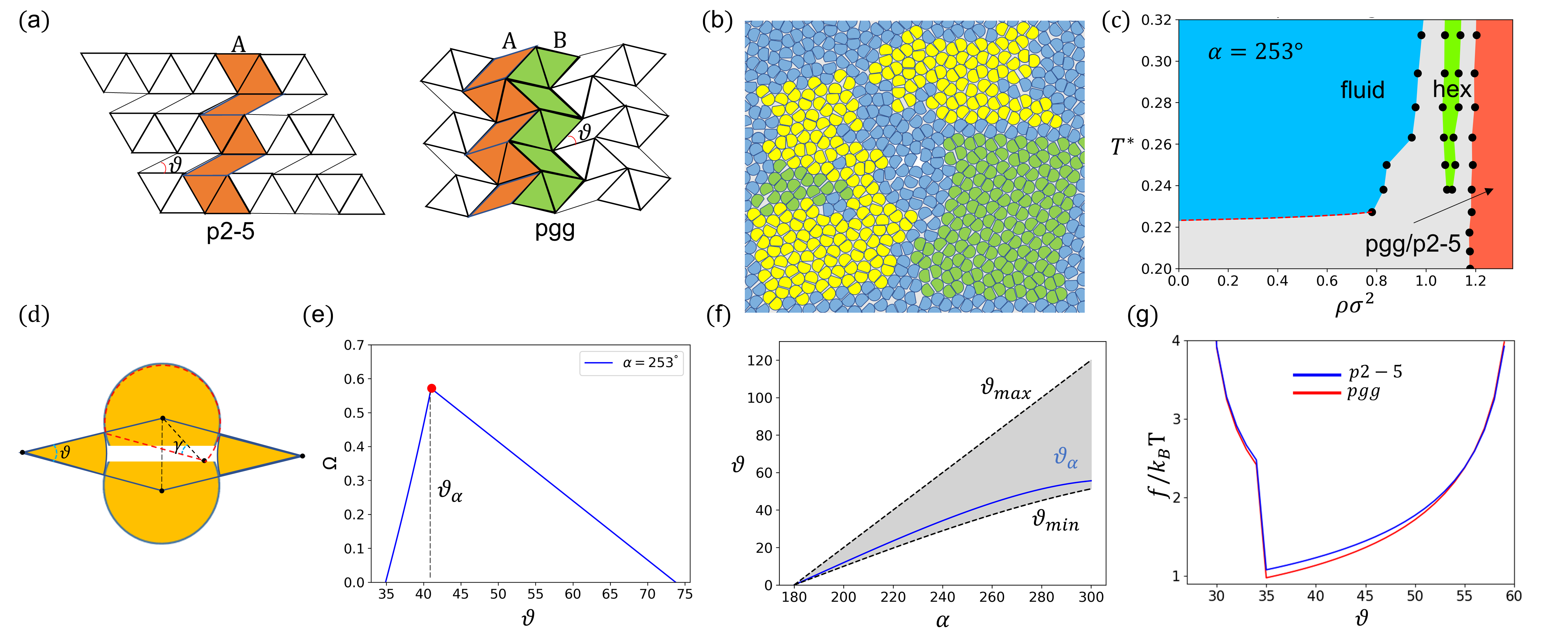} }
	\caption{  (\textbf{a})  Columns $A$ and $B$ are building blocks of \emph{p2-5} and \emph{pgg}. (\textbf{b}) Simulation snapshot at $\alpha=253^\circ$, $T^*= 0.227 $ and  $P=6 \varepsilon/\sigma^2$, where \emph{p2-5}  and \emph{pgg} structures are marked in green and yellow colors, respectively. (\textbf{c})  Phase diagram of the system in dimensions of  $\rho$ and  $T^*$ at $\alpha=253^\circ$. (\textbf{d})  Definitions of angles $\gamma$ and $\vartheta$. The red dashed lines represent the fluctuating state of particles. (\textbf{e}) Rotational microscopic states $\Omega$ for both two crystals as a function of $\vartheta$. (\textbf{f}) $\vartheta_{\alpha}$, $\vartheta_{max}$ and $\vartheta_{min}$  as a function of  $\alpha$. (\textbf{g}) Free energy per particle for \emph{p2-5} and \emph{pgg} harmonic networks as a function of $\vartheta$ at  $T^*=0.2$. \label{stability}  }
\end{figure*}

The \emph{pgg} and \emph{p2-5} structures are {previously unreported} self-assembled structures with coordination number $z=5$. In Figure~\ref{stability}a, we show that  \emph{p2-5}  structures can be viewed as the stacking of the parallel columns $A$, while \emph{pgg} structure is a result of the stacking of parallel columns $A$ and  $B$ alternatively. Here, $B$ is the mirror reflection of $A$. Therefore, the deformation of  \emph{p2-5} and \emph{pgg} structures are both governed by the open angle $\vartheta$, as  defined in Figure~\ref{stability}a. It can be seen that for the \emph{pgg} structure, $\vartheta$ in all cells change synchronously since \emph{pgg} structure has only one uniform soft mode. On the contrary, the \emph{p2-5} structure lacks uniform soft modes but exhibits multiple horizontal soft modes owing to the parallel rigid columns \cite{deng2020supercrystallographic} stacking in the top-down direction.

 In the simulation of the system at $\alpha=253^\circ$, we observe the nucleation of both \emph{p2-5} and \emph{pgg} crystal phases from a disordered fluid phase, as depicted in Figure~\ref{stability}b. This suggests that free energies of these two crystals at the same pressure are very closed, analogous to that of FCC and HCP crystals of hard spheres, as well as structures previously reported in Ref.~\cite{lei2021self}.  Since the ground state energies of the \emph{p2-5} and \emph{pgg} structures are the same,  entropy plays the dominant role in determining which structure is more stable at finite temperature. The entropy of the self-assembled structures are mainly composed of the rotational entropy and vibrational entropy of the particles. As shown in Figure~\ref{stability}d, in both the  \emph{p2-5} and \emph{pgg} structures, the rotation of particles are highly constrained by the coordination bonds and excluded volume of other particles. Under mean-field approximation (see Section A and Figure~S3 in SI for details), we can obtain the allowed rotational free space (or angle) $\Omega$ per particle as a function of $\vartheta$,
\begin{eqnarray}
\Omega =\left\{
\begin{array}{ll}
2\arcsin(4\sin\frac{\vartheta}{2}-\sin\gamma)-2\gamma &   ~~ \vartheta \leq \vartheta_\alpha,  \\
2\gamma-\vartheta  & ~~ \vartheta \geq \vartheta_\alpha,
\end{array}\right.             \label{Omega}
\end{eqnarray}
where $\gamma=(\alpha-\pi)/2$, and $\vartheta_\alpha$ is the equilibrated open angle at which $\Omega$ or the rotational entropy is maximized,
\begin{eqnarray}
\vartheta_\alpha = 2\arcsin \left( \frac{\sin\gamma}{A} \right) + 2\arctan \left(\frac{\sin 2\gamma}{4+\cos 2\gamma} \right)
\end{eqnarray}
with $A=\sqrt{(4+\cos 2\gamma)^2+(\sin 2\gamma)^2}$. This relationship is the same for both the \emph{p2-5} and \emph{pgg} structures. This means that at the mean-field level, particles in the two structures have the same rotational entropy  $S_{\rm rot}=k_B \ln \Omega$. In Figure~\ref{stability}e, we show $\Omega$ as a function of $\vartheta$ for the system under $\alpha=253^\circ$. It is evident that the rotational entropy is maximized at $\vartheta_{\alpha}$. This indicates that $\vartheta_{\alpha}$ is the thermodynamically stable open angle at finite temperature. Furthermore, in Figure~\ref{stability}f, we depict $\vartheta_{\alpha}$ as a function of $\alpha$. We observe a monotonic increase in $\vartheta_{\alpha}$ before it reaches a saturation point around $60^\circ$. The maximum and minimum values of $\vartheta$, denoted as $\vartheta_{max}$ and $\vartheta_{min}$ respectively, are also represented by dashed lines. These lines define the range (shown as a shadow area) within which the structures can be stretched or compressed without altering the coordination.

When $w \rightarrow 0$ and $T^{*} \rightarrow 0$, the rods can only do vibration motion around the equilibrium position. The system thus can be described by an effective Hamiltonian under the harmonic approximation,
\begin{equation}
H_{\mathrm{eff}}=\sum_{i j} \frac{k}{2} [(\mathbf{u}_{i}-\mathbf{u}_{j})\cdot \mathbf{e}_{ij}]^{2}+\frac{\kappa}{2} \sum_{i} \Delta \alpha_{i}^{2}  \label{Hamilonian}
\end{equation}
where the first summation represents the energy contribution from central-force bonds with $\mathbf{u}_{i}=\mathbf{r}_{i}-\mathbf{R}_{i}$ the translational displacement of particle $i$ from its equilibrium position $\mathbf{R}_{i}$, and  $\mathbf{e}_{ij}$  is  the normalized bond vector. The second summation represents the contribution of bond-bending rigidity, where $\Delta \alpha_{i}=\alpha_{i}-\alpha_{s}$ represents the deviation of the bond angle $\alpha_i$ from its equilibrated value, which is dependent on $\mathbf{u}_{i}$, as explained in the Method section. This rigidity arises from the rotational entropy of the particles and is an entropic effect. The effective spring constant $k$ for the central-force bonds can be approximated as $k \simeq k_{B} T / w^{2}$,  based on the energy partition of an oscillator. The coefficient $\kappa$ represents the bond-bending rigidity and its expression can be found in the Method section. The above Hamiltonian can be written in a more compact form, namely,
\begin{equation}
H_{\mathrm{eff}}=\frac{1}{2} \sum_{l, l^{\prime}} \overrightarrow{\mathbf {U}}_{l} \cdot \mathbf{D}_{l, l^{\prime}} \cdot \overrightarrow{\mathbf{U}}_{l^{\prime}}
\end{equation}
with $\overrightarrow{\mathbf{U}}_{l}=\sum_{p} \overrightarrow{\mathbf{u}}_{l}^{p}$. Here, $\overrightarrow{\mathbf{u}}_{l}^{p}$ is the 4$\times$2 dimension displacement vector of particle $p$ in unite cell $l$. $\mathbf{D}_{l, l^{\prime}}$ is the dynamic matrix of the lattice (see the Method section for details). It should be noted that  although \emph{p2-5} and \emph{pgg} structures share the same $k$ and $\kappa$, due to the different arrangement of the force bonds, their dynamic matrices $\mathbf{D}_{l, l^{\prime}}$ are different.

\begin{figure*}[htbp] 
	\resizebox{180mm}{!}{\includegraphics[trim=0.0in 0.0in 0.0in 0.0in]{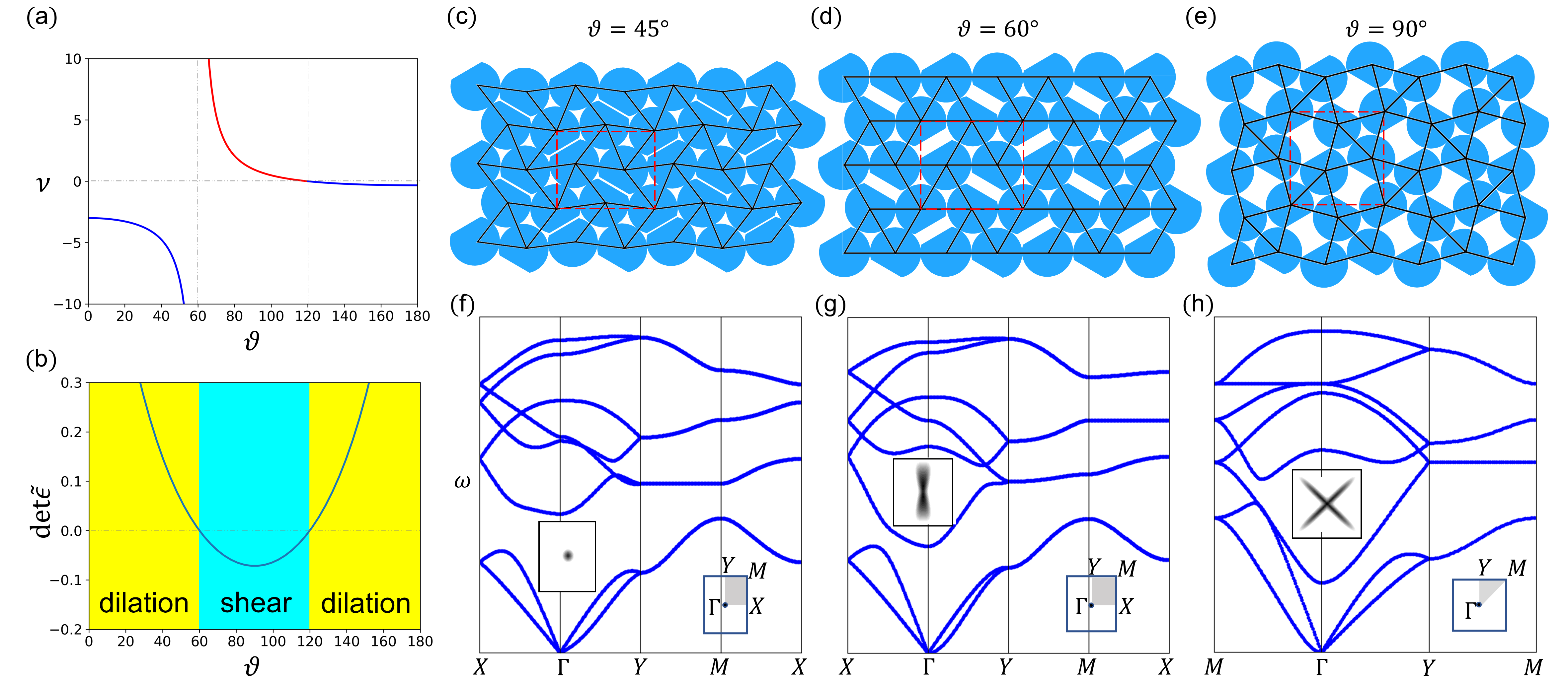} }
	\caption{  (\textbf{a}) The Poisson's ratio of the \emph{pgg} structure. (\textbf{b}) $\det \tilde{\epsilon}$ of the \emph{pgg} structure. (\textbf{c-e}) \emph{pgg} structures at $\vartheta=45^\circ,60^\circ,90^\circ$, respectively. (\textbf{f-h}) The corresponding bulk phononic spectrum for  \emph{pgg} structures at $\vartheta=45^\circ,60^\circ,90^\circ$, respectively.  Inset is the density plot of the lowest branch  phononic spectrum. }  \label{bands}
\end{figure*} \textbf{}

 With the dynamic matrix, the free energy of the harmonic networks can be calculated based on the Gaussian integration,~\cite{mao2013entropic,lei2021self}
\begin{equation}
f=\frac{k_{B} T}{2 m} \int_{1 \mathrm{BZ}} \frac{v_{0} {\rm d} \mathbf{q}}{(2 \pi)^{d}} \ln \operatorname{\det} \tilde{\mathbf{D}}_{\mathbf{q}}-\frac{d}{2} k_{B} T \ln \left(\frac{2 \pi \Lambda^{2}}{\sigma^{2}}\right),
\end{equation}
where $ \tilde{\mathbf{D}}_{\mathbf{q}} $ is the normalized dynamic matrix in $q$ space  and $v_0$ is the volume of unit cell.  The integration is over the first Brillouin zone (BZ).  The free energies per particle of \emph{p2-5} and \emph{pgg} harmonic crystals as a function of $\vartheta$ at $T^* \simeq 0.1$ are shown in Figure~\ref{stability}g, which indicates the free energy of the \emph{pgg} crystal is slightly lower than that of the \emph{p2-5} crystal. But at other temperatures, we find it can be the opposite (see Figure~S4 in SI).  According to the equipartition theorem, the energy of \emph{p2-5} and \emph{pgg} harmonic networks are the same under the same temperature. Therefore, the difference in free energy primarily arises from the collective vibration entropy of the particles. In practice, since the free energy difference between these two crystals is small, one can use prefabricated template to induce the nucleation of the preferred crystal phase.~\cite{hynninen2007self,van1997template} In our simulation, we use 2D tiles to mimic 3D colloidal rods, neglecting the fluctuation of rods’ orientations. Explicitly including this fluctuation is not expected to fundamentally change the phase behaviors of the systems, as the ground-states of the 2D and 3D systems, i.e., the crystal phases, are the same.

\subsection*{Soft Modes in \emph{pgg} Structures}
In order to explore the mechanical properties of the \emph{pgg} structure, we consider the scenario where the bond-bending rigidity of the crystal in Eq.~(\ref{Hamilonian}) can be neglected, i.e., $k \sigma^{2} \gg \kappa$. This can be  realized by decreasing either temperature $T^*$ or potential well width $w$. In this limit, the \emph{pgg} structure would be a standard super-isostatic lattice with an uniform soft mode, which can be represented by a strain tensor in the regular coordination, i.e., coordination that matches  the rectangular unit cell,~\cite{rocklin2017transformable}
\begin{equation}
\tilde \epsilon
=\begin{pmatrix}
\tilde{\epsilon}_{x x} & \tilde{\epsilon}_{x y} \\
\tilde{\epsilon}_{x y} & \tilde{\epsilon}_{y y}
\end{pmatrix}
=\frac{1}{2}\begin{pmatrix}
  - \tan(\frac{\vartheta}{2}-\frac{\pi}{6}) &0 \\
  0& \cot (\frac{\vartheta}{2}+\frac{\pi}{6})
\end{pmatrix}  \delta \vartheta  \label{uniform_strain}
\end{equation} 
Since the off-diagonal of the matrix is zero,  the uniform soft mode only changes the aspect ratio of the unit cell. The  Poisson ratio of the pgg structure under this coordination can be calculated as,
\begin{align}
v=-\frac{\tilde{\epsilon}_{y y} }{\tilde{\epsilon}_{x x} }=\cot\left(\frac{\vartheta}{2}+\frac{\pi}{6}\right)\cot\left(\frac{\vartheta}{2}-\frac{\pi}{6}\right).
\end{align}
In Figure~\ref{bands}a, we plot Poission ratio $v$ as a function of $\vartheta$. We find that in the range of $60^\circ \sim 120^\circ$,  $v$ is positive, while in the range of  $0^\circ \sim 60^\circ$ and $120^\circ \sim 180^\circ$, $v$ is negative (auxetic).~\cite{greaves2011poisson}  Similar results were also reported in Ref.~\cite{czajkowski2022duality}. It should be emphasized that structures with $\vartheta$ and $180^\circ- \vartheta$ are in fact the same structure connected by $90^\circ$ rotation due to the symmetry of this structure.

In Ref.~\cite{rocklin2017transformable}, by using continuous elastic medium (CEM) theory,  Rocklin et al. proven that when a 2D elastic system has an uniform soft mode, there must exists two families of spatially-varying soft modes (also see Section B in SI for the detailed derivation),
\begin{align}
\begin{array}{l}
\tilde{\epsilon}_{+}(\mathbf{r})=\tilde{\epsilon}~ \psi_{+}\left(x+\lambda_{+} y\right), \\
\tilde{\epsilon}_{-}(\mathbf{r})=\tilde{\epsilon}~ \psi_{-}\left(x+\lambda_{-} y\right),
\end{array}
\end{align}
where $\psi \pm(s)$ are two arbitrary scalar functions of $s$ and 
\begin{align}
\lambda_{\pm}=\left(\tilde{\epsilon}_{x y} \pm \sqrt{-\det \tilde{\epsilon}}\right) / \tilde{\epsilon}_{x x}
\end{align}
where $\det \tilde{\epsilon}=\tilde{\epsilon}_{x x} \tilde{\epsilon}_{y y}-\left(\tilde{\epsilon}_{x y}\right)^{2}$ is the determinant of $\tilde{\epsilon}$, which is independent of the choice of coordinates. It is obvious from the formula that whether $\lambda_{\pm}$ is real or complex depends on the sign of $\det \tilde{\epsilon}$, based on which  two regimes emerge, i.e., the shear dominant regime with $\det \tilde{\epsilon}<0$ and dilation dominant regime  with $\det \tilde{\epsilon} >0$ (Figure~\ref{bands}b).  For the \emph{pgg} structure, the determinant of $\tilde{\epsilon}$ is found to have opposite sign to the Poisson ratio. Therefore, the shear and dilation regimes correspond precisely to the positive and negative Poisson ratio regimes, respectively. It can be proven that soft modes with real $\lambda_{\pm}$ in the shear dominant regime are bulk plane waves along two coordination-independent directions 
\begin{eqnarray}
\boldsymbol{a}_1 = (1,  \lambda_+ ) ~~~ {\rm and} ~~~  \boldsymbol{a}_2 = (1,  \lambda_-  ),
\end{eqnarray}
while soft modes with imaginary $\lambda_{\pm}$ in the dilation dominant regime are edge modes along the direction of  $ (1, 0)$, which is coordination dependent. This indicates these edge modes exist on all boundaries. Recently, a  theoretical framework based on complex analyticity has been developed from CEM theory.~\cite{czajkowski2022duality}

\begin{figure*}[htbp] 
	\resizebox{140mm}{!}{\includegraphics[trim=0.5in 0.0in 0.0in 0.0in]{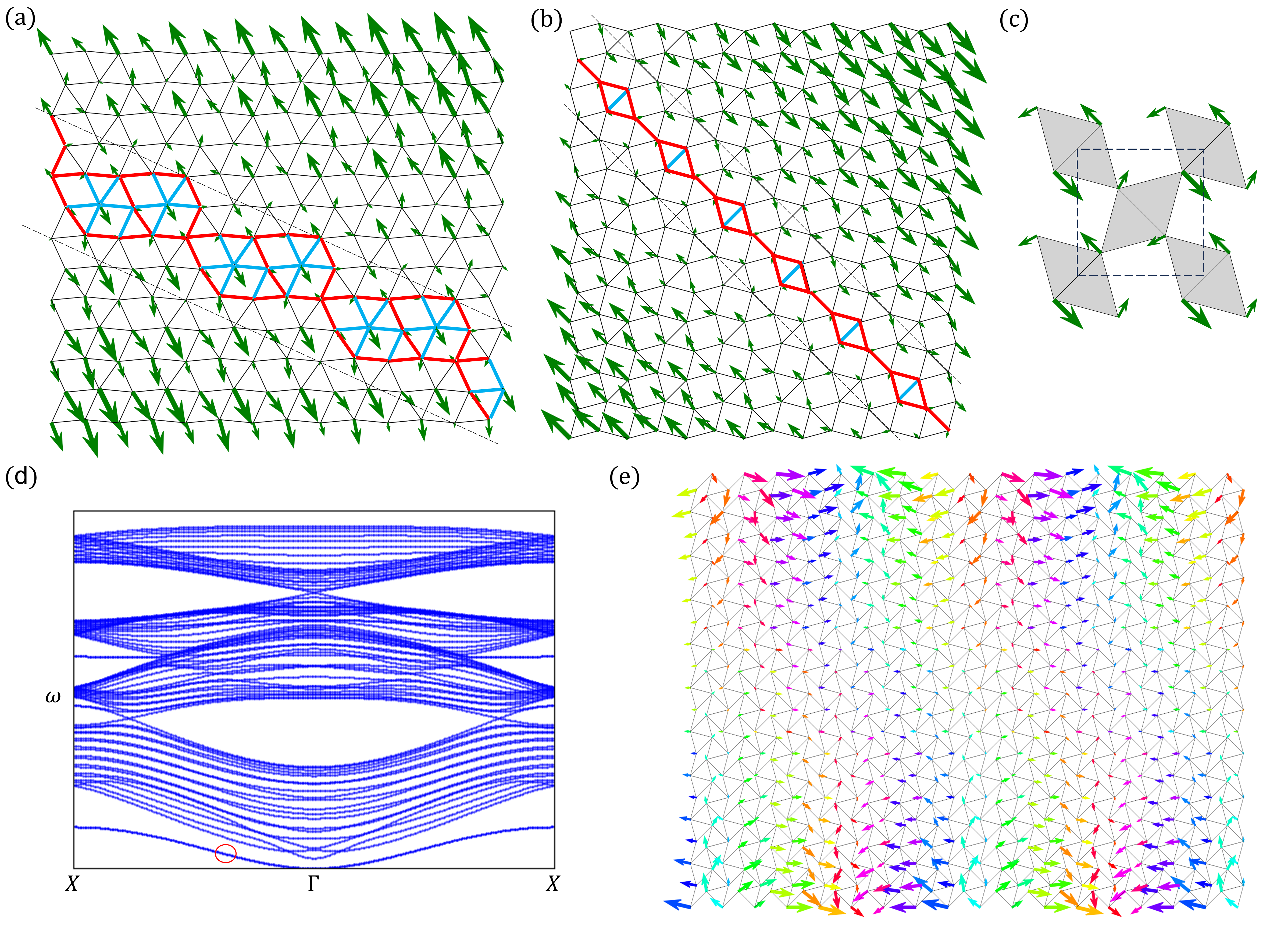} }
	\caption{ (\textbf{a}) Soft bulk modes of the \emph{pgg} structure at $\vartheta=70.9^\circ$, where $q_x=0.03\sigma$, $q_y=0.0657\sigma$.  (\textbf{b}) Soft bulk modes at $\vartheta=90^\circ$, where $q_x=q_y=-0.05\sigma$. The two black dash lines enclose a SSS, where the bonds under tension or compression are drawn with red or blue colors, respectively.  (\textbf{c}) The relative vibration displacement of particles with respective to the center of the unit cell for (\textbf{b}), which approximates the uniform soft mode. (\textbf{d}) The phononic spectrum for the system with $\vartheta=30^\circ$  under open boundary condition in $y$-direction. (\textbf{e}) Illumination of the edge modes, where the arrow size and color represent the instantaneous particle vibration displacement and phase, respectively.  This edge mode is marked by red circle in  (\textbf{d}). } \label{edge mode}
\end{figure*}

In order to verify the above CEM theory, we calculate the phononic spectrum for pgg crystals with  different open angle $\vartheta$ under periodic boundary conditions (PBCs).  We confirm that the soft bulk modes only exist in the shear dominant regime $60^\circ \le \vartheta \le 120^\circ$ and the soft mode direction is along $\boldsymbol{a}_1 $ and $\boldsymbol{a}_2$, fully agreeing with the theoretical predictions.  Typical results for $\vartheta=45^\circ,60^\circ,90^\circ$  are shown in Figure~\ref{bands}, where the 2D surfaces of the lowest frequency branches are drawn in gray level in the insets. It can be seen that when $\vartheta=90^\circ$, there are two non-linear bulk soft modes along the diagonal direction near the $\Gamma$ point (inset of Figure~\ref{bands}h). When $\vartheta$ approaches to $60^\circ$, These two bulk soft modes merge. When $\vartheta$ further decreases, all the dispersion relationships become linear as shown by the case of $\vartheta=45^\circ$ (Figure~\ref{bands}f). This is accompanied by the disappearance of the bulk soft modes and the occurrence of edge soft modes on all boundaries.


To investigate the mechanism behind the non-linear dispersion relationship around the $\Gamma$ point, we utilize the observation that the determinant of ${\tilde{\mathbf{D}}}(\boldsymbol{q})$, which represents the product of all eigenvalues, is governed by the acoustic modes near zero frequency in the vicinity of the $\Gamma$ point.~\cite{kane2014topological}Therefore one can obtain the hybrid dispersion relationship for the acoustic bands near  $\Gamma$ points  by expanding $ {\tilde{\mathbf{D}}}(\boldsymbol{q})$ and keep only the lowest order terms, namely,
\begin{equation}
\omega_1^2 \omega_2^2(\boldsymbol{q}) \propto \det  {\tilde{\mathbf{D}}}(\boldsymbol{q})=C_1 q_x^4 + C_2 q_y^4 +C_3 q_x^2q_y^2 +O(\boldsymbol{q}^6)
\end{equation}
where $\omega_1$ and $\omega_2$ are the two lowest  eigen-frequency of  acoustic bands. $C_1,C_2,C_3$ are coefficients that depend on $\vartheta$ (see Section C in SI for details). For mechanical stable structures with linear dispersion relationships, as least one of these  fourth order terms is non-zero. Non-linear quadratic dispersion emerges when linear terms vanish,
\begin{equation}
C_1 q_x^4 + C_2 q_y^4 +C_3 q_x^2q_y^2 =0.
\end{equation}
Solving this equation results in wave vector constrained along two directions
\begin{equation}
q_y/q_x = \lambda_{\pm} 
\end{equation}
which is exactly the soft mode directions predicted by the CEM theory. In these two directions,  we have $\omega_1^2 \omega_2^2(\boldsymbol{q}) \propto  O(\boldsymbol{q}^6)$. For this hybrid dispersion relationship, if one mode remains linear dispersed, the other mode must has a quadratic dispersion relationship $\omega \propto   q^2$, which is much softer than the linear mode. These predictions are  confirmed by our phononic spectrums as shown in Figure~\ref{bands}f-h.

In addition to the explanation provided above, there is another more intuitive interpretation for the mechanism of bulk soft mode. First,  one can prove that the divergence (or vanishing)  of the Poisson ratio of the \emph{pgg} structure at $\vartheta=60^\circ$  (or $\vartheta=120^\circ$) in regular coordination is accompanied by the zero strain under uniform soft deformation in the $x$ (or $y$) direction. {More generally, the strain $\tilde{\epsilon}_{\boldsymbol{k}}$ in an arbitrary direction ${\boldsymbol{k}} = (\cos\phi, ~\sin\phi) $ due to the uniform soft deformation can be written as,
\begin{equation}
\tilde{\epsilon}_{\boldsymbol{k}} = \frac{1}{2} \left[ \frac{\sin \phi \tan \phi }{\tan \left(\frac{\vartheta}{2}+\frac{\pi}{6}\right)} -  \cos \phi \tan \left(\frac{\vartheta}{2}-\frac{\pi}{6}\right) \right] \delta \vartheta.
\end{equation} 
For $60^\circ<\vartheta<120^\circ$, $\tilde{\epsilon}_{\boldsymbol{k}}$ can be zero during the uniform soft deformation at two special directions in the regular coordination,
\begin{eqnarray}
\boldsymbol{b}_1 = (-\lambda_+, 1) ~~~ {\rm and} ~~~ \boldsymbol{b}_2 = (-\lambda_-, 1).
\end{eqnarray}
Importantly, the zero strain in certain directions during the uniform soft deformation indicates the existence of the states of self stress (SSS) along $\boldsymbol{b}_1$ and $\boldsymbol{b}_2$ under PBCs. These SSSs are not necessary associated with straight lines of bonds as previously reported.~\cite{rocklin2017transformable,zhou2018topological}  According to index theorem,~\cite{calladine1978buckminster,lubensky2015phonons} every SSS is associated with an additional soft mode. These soft modes are usually along the direction of SSSs line direction. In periodic structures, there are infinite  SSS lines arranged in parallel. The corresponding soft modes are the transverse wave modes perpendicular to the SSS lines.  Indeed, one can check that $\boldsymbol{b}_1$ and $\boldsymbol{b}_2$ are perpendicular to $\boldsymbol{a}_1$ and $\boldsymbol{a}_2$ respectively, which indicates that these SSSs underlie the soft bulk modes observed in the shear dominant regime. To confirm this prediction, in Figure~\ref{edge mode}a,b, we show one soft bulk mode under $\vartheta=70.9^\circ$ ($q_x = 0.03\sigma$, $q_y = 0.0657 \sigma $) and $\vartheta=90^\circ$ ($q_x = q_y = -0.05 \sigma $) where the arrows represent the eigenmode of vibration.  The  SSSs  represented by the bonds under tension or compression are drawn with red or blue colors, respectively. It can be observed that bulk soft modes are indeed the transverse mode perpendicular to the SSS direction. Additionally, by visually representing the relative infinitesimal displacement of particles within the unit cell, we obtain the eigenmode of the uniform soft mode (Figure\ref{edge mode}c). This analysis unveils that bulk soft modes at low $q$ values are  a spatial modulation of the uniform soft mode, which aligns with previous CEM  theory.

In the dilation dominant regime ($0^\circ < \vartheta < 60^\circ$ and $120^\circ  < \vartheta < 180^\circ$), the strain under uniform soft deformation cannot be zero in arbitrary direction, excluding the existence of bulk soft modes induced by SSSs in PBCs. Nevertheless, in the open boundary conditions (OBCs), boundary cutting reduces the number of constraint by the order of $\sqrt{N}$, creating equal number of soft modes. These modes are not bulk soft modes (otherwise also exist in PBCs). Therefore, they must be localized at the surface. In the shear dominant regime, the soft bulk modes would exist in OBCs, but no additional soft edge modes are created by OBCs.  This is because boundary cutting eliminates the SSSs. Thus, the number of removed constraint equals  the number of reduced SSSs.  Based on the  index theorem, the total number of soft modes remains unchanged. To study the edge modes in the dilation dominant regime more explicitly, we calculate the phononic spectrum of the system with ten primitive cells arranged in the y-direction (OBCs) and PBC in the $x$-direction  under  $\vartheta=30^\circ$ (Figure~\ref{edge mode}d). We find the lowest band dropping to zero frequency at $\Gamma$ point for $\vartheta<60^\circ$ or $\vartheta>120^\circ$. In Figure~\ref{edge mode}e, we show the eigenmodes of this band and find that it decays quickly from the surface to the bulk, a typical behaviour of soft edge modes. 

In the above, we have demonstrated that the \emph{pgg} structures with different $\vartheta$ can act as two distinct types of super-isostatic structures. One type exhibits soft bulk modes and rigid edge modes, while the other exhibits soft edge modes and rigid bulk modes. Notably, $\vartheta$ can be smoothly adjusted without any energy cost, making the \emph{pgg} structures transformable super-isostatic structures.~\cite{sun2012surface,liarte2020multifunctional,czajkowski2022duality} Examining the self-assembled \emph{pgg} structures in Figure~\ref{stability}f, we observe that when $\alpha$ is approximately $260^\circ$, the permissible range of $\vartheta$ spans from $40^\circ$ to $80^\circ$, encompassing the transition point at $\vartheta = 60^\circ$. Consequently, the self-assembled structures are capable of transforming between two different soft states through stretching or compression. The resistant force required during this transformation is the entropy force, which is relatively small for colloidal crystals at low temperatures and can be  further reduced by modifying the attractive area on the particles.~\cite{edmond2021large,bahri2022self}

\section*{Conclusion}
In this work, by using Monte Carlo simulation, we  systematically study the self-assembly of perfectly-aligned segment rods with lateral cutting and anisotropic short range attraction. We find that rods with different cutting degrees can self-assemble into different ordered structures with average coordination $z$ varying from 3 to 6. We especially  focus on the  self-assembled \emph{pgg} crystals with  $z=5$. This structure is a transformable super-isostatic structure capable of uniform soft deformation. We find that the existence of the uniform deformation modes  underlie the non-uniform soft bulk modes at open angle  $60^\circ<\vartheta<90^\circ$ and soft edge modes at $0^\circ<\vartheta<60^\circ$. The transformation between the bulk and edge soft states can be realized through the uniform soft deformation. We prove that the bulk soft mode are non-linear quadratic modes.  Importantly, we show that these soft bulk modes are associated with SSSs in the zero strain direction during the uniform soft deformation, which provides a {more intuitive} interpretation for the origin of soft bulk mode. Our work provides a feasible strategy to fabricate  transformable phononic/mechanical meta-materials at micro-scales, whose properties can be effectively switched  through compression or stretching. We also expect that other types of colloidal particles, like hemi-sphere or bowl-shape particles,~\cite{lei2021self,edmond2021large} can be self-assembled into 3D isostatic or super-isostatic transformable structures, which will be a focus of our further works.

\section*{Methods}
\subsection*{Particle Rotational Entropy  and Bond Angle Rigidity}
In the low temperature limit, the energies of \emph{p2-5} and \emph{pgg} structures are the same due to the identical coordination number. The stability of  two structures then are solely determined by the entropy, which is composed of rotational  and vibrational entropies of particles. The rotational entropy of a particle is determined by the allowed angle space $\Omega$ that the particle can freely rotate, which is restricted by the cutting angle of the particle and the confinement from adjacent  particles. $\Omega$  is calculated at the mean-field level, i.e., assuming that  neighbour particles take the  equilibrated position. To obtain  bond angle rigidity $\kappa$ at arbitrary open angle, we further suppose that the crystal network is maintained by external force at $\vartheta=\vartheta_0$ and assume the symmetry bond angle fluctuation $\Delta \vartheta$ around the equilibrated $\vartheta_0$. Under harmonic approximation, the effective Hamiltonian per particle for the bond angle bending can be written as
\begin{equation}
	H^{\text {angle }} \simeq \frac{\kappa}{2}(\Delta \vartheta)^{2}
\end{equation}
Detailed calculation can be found in the Section A of SI.

\subsection*{Derivation of Dynamic Matrix}
As a generalization of Ref.~\cite{mao2011coherent}, We first consider a crystal lattice with $n$ particles in a primitive unit cell. We construct a $n$-dimensional vector $ \hat{\mathbf{P}}$  with each dimension representing the number of particle,
\begin{eqnarray}
{\mathbf {\hat P} }_{1}=
\begin{pmatrix}
1\\ 0 \\ 0\\ \vdots
\end{pmatrix}
,
{\mathbf {\hat P} }_{2}=
\begin{pmatrix}
0\\ 1 \\ 0 \\ \vdots
\end{pmatrix}
,
\cdots,~~~
{\mathbf {\hat P} }_{n}=
\begin{pmatrix}
\vdots  \\ 0 \\ 0  \\ 1 
\end{pmatrix}
\end{eqnarray}
which satisfies the orthogonality $ \hat{\mathbf{P}_i} \cdot  \hat{\mathbf{P}_j}=\delta_{ij}$. We define the $d$ dimension  displacement vector ${\mathbf u}^{p}_{l}$, where  $l$ is the index of cell and $p$ is the index of the particle in the cell.  To construct the dynamic matrix, we introduce a $nd$ dimension vector to represent the displacement of particles $p$ in the cell $l$, i.e.,
\begin{eqnarray}
\overrightarrow{ {\mathbf u}}^{p}_{l}  =  {\mathbf u}^{p}_{l} \otimes  {\mathbf {\hat P} }_{p} 
\end{eqnarray}
where $\otimes$ is the Kronecker product. Similarly, the bond vector pointing from particle $(p_1,l_1)$ to particle $(p_2,l_2)$ in $nd$-dimensional space can be written as 
\begin{eqnarray}
\overrightarrow{ {\mathbf e}}_{l_{1}l_{2}}^{p_{1}p_{2}}
&=
(1-\delta_{p_{1}p_{2}}) {\mathbf e}_{l_{1}l_{2}}^{p_{1}p_{2}} \otimes
{\mathbf {\hat P} }_{p_{1}} + {\mathbf e}_{l_{1}l_{2}}^{p_{1}p_{2}} \otimes {\mathbf {\hat P} }_{p_{2}}
\end{eqnarray}
where $\mathbf{e}_{l_{1} l_{2}}^{p_{1} p_{2}}$ is the corresponding bond vector in $d$-dimension space. We can prove that $\overrightarrow{\mathbf{e}}_{l_{1} l_{2}}^{p_{1} p_{2}}$ satisfies
\begin{eqnarray}
\overrightarrow{ {\mathbf e}}_{l_{1}l_{2}}^{p_{1}p_{2}}  \cdot \overrightarrow{ {\mathbf u}}^{p}_{l}
&= { {\mathbf e}_{l_{1}l_{2}}^{p_{1}p_{2}} } \cdot  {\mathbf u}^{p}_{l} ~~~~~(p=p_1,p_2)
\end{eqnarray}
especially when $p_1=p_2$. The  effective Hamiltonian of the central-force harmonic bond network can be written as,
\begin{eqnarray}
	H_{\rm bond} =  \frac{1}{2} \sum_{l,l'}  \sum_{p,p'}   \overrightarrow{\mathbf u}_{l}^{p}  \cdot  {\mathbf D}^{\rm bond}_{l,l'}  \cdot  \overrightarrow{\mathbf u}_{l'}^{p'}
\end{eqnarray}
with 
\begin{eqnarray}
\begin{split}
	{\mathbf D}^{\rm bond}_{l,l'}  =   \sum_{b} {k}     
	\left[ { {\mathbf e}}_{l_{1}l_{2}}^{p_{1}p_{2}}    \otimes \left(   {\mathbf {\hat P} }_{p_1} \delta_{l,l_{1}} 
	-   {\mathbf {\hat P} }_{p_2}  \delta_{l,l_{2}}      \right)  \right] 
	\\ \times \left[ { {\mathbf e}}_{l_{1}l_{2}}^{p_{1}p_{2}}   \otimes 
	\left(   {\mathbf {\hat P} }_{p_1} \delta_{l',l_{1}} 
	-    {\mathbf {\hat P} }_{p_2}  \delta_{l',l_{2}}      \right)  \right].
\end{split}
\end{eqnarray}
the $nd\times nd$ dynamic matrix for central-force bond interaction. Here $\sum_{b}$ represents the sum over all bonds in the system. In Fourier space, the  dynamic matrix can be written as  
\begin{eqnarray}
{\mathbf D}^{\rm bond}_{ {\mathbf q}, {\mathbf  q'}} = N_c \delta_{{\mathbf q} ,{\mathbf q} '} {\mathbf D}^{\rm bond}_{\mathbf q}  \label{D_bond_qq}
\end{eqnarray}
with $N_c$ the number of cell and
\begin{eqnarray}
{\mathbf D}^{\rm bond}_{\mathbf q} &=& k  \sum_{s}     \overrightarrow{ \mathbf b }^b_{s,{\mathbf q} }  \overrightarrow{ \mathbf b }^b_{s,-{\mathbf q} }   \label{D_bond}   \\
\overrightarrow{ \mathbf b }^b_{s,{\mathbf q} } &=& 
{\mathbf e}_{l^s_{1}l^s_{2}}^{p^s_{1}p^s_{2}} \otimes \left[ 
{\mathbf {\hat P} }_{p^s_{1}}  - e^{-i{\mathbf q} \cdot \left({\mathbf r} _{l^s_{2}}-{\mathbf r}_{l^s_{1}} \right)} 
{\mathbf {\hat P} }_{p^s_{2}} \right].
\end{eqnarray}
Here $\sum_s$ contain all bonds that connect particles in a single unit cell with $s$ the index of bond. $r_l$ is the center of unit cell $l$.  It should be mentioned that the matrix formed by $\overrightarrow{ \mathbf b }^b_{s,{\mathbf q} }$ arranged in columns is the compatibility matrix ${\mathbf C}$.~ \cite{sun2012surface} To obtain ${\mathbf D}^{\rm bond}_{\mathbf q} $ for \textit{pgg} crystal, we first define angle
\begin{equation}
\widetilde{\alpha}=\frac{\pi}{6}-\frac{\vartheta}{2}
\end{equation}
and  two primitive vectors
\begin{align}
\mathbf{a_1} &=[\ 2\cos\widetilde{\alpha},\ 0\ ] \\
\mathbf{a_2} &=[\ 0,\ 2\sin\left(\frac{\pi}{3}-\widetilde{\alpha}\right)\ ].
\end{align}
The coordinations for the 4 particles in the unit cell of pgg lattice are
\begin{align}
\mathbf{r_1} &= [\ 0,\ 0\ ] \\
\mathbf{r_2} &= [\ \cos\widetilde{\alpha},\ -\sin\widetilde{\alpha}\ ] \\
\mathbf{r_3} &= [\ \frac{3}{2}\cos\widetilde{\alpha}+\frac{\sqrt3}{2}\sin\widetilde{\alpha},\ \frac{\sqrt3}{2}\cos\widetilde{\alpha}-\frac{3}{2}\sin\widetilde{\alpha}\ ] \\
\mathbf{r_4} &= [\ \frac{1}{2}\cos\widetilde{\alpha}+\frac{\sqrt3}{2}\sin\widetilde{\alpha},\ \frac{\sqrt3}{2}\cos\widetilde{\alpha}-\frac{1}{2}\sin\widetilde{\alpha}\ ].
\end{align}
By defining the  bond vector as ${\mathbf e}_{ij} = {\mathbf r}_j - {\mathbf r}_i $, we can have
\begin{equation}
\begin{aligned}
\overrightarrow{\mathbf{b}}^b_{1,q} &= [\ \mathbf{e}_{12},\ -\mathbf{e}_{12},\ 0,\ 0\ ]  \\
\overrightarrow{\mathbf{b}}^b_{2,q} &= [\ \mathbf{e}_{14},\ 0,\ 0,\ -\mathbf{e}_{14}\ ]  \\
\overrightarrow{\mathbf{b}}^b_{3,q} &= [\ 0,\ \mathbf{e}_{24},\ 0,\ -\mathbf{e}_{24}\ ]  \\
\overrightarrow{\mathbf{b}}^b_{4,q} &= [\ 0,\ \mathbf{e}_{23},\ -\mathbf{e}_{23},\ 0\ ]  \\
\overrightarrow{\mathbf{b}}^b_{5,q} &= [\ 0,\ 0,\ \mathbf{e}_{34},\ -\mathbf{e}_{34}\ ]  \\
\overrightarrow{\mathbf{b}}^b_{6,q} &=(\mathbf{e}_{21}+\mathbf{a}_{1}) [\ -e^{-i\mathbf{q}\cdot\mathbf{a}_1},\ 1,\ 0,\ 0\ ]  \\
\overrightarrow{\mathbf{b}}^b_{7,q} &=(\mathbf{e}_{43}-\mathbf{a}_{1}) [\ 0,\ 0,\ -e^{i\mathbf{q}\cdot\mathbf{a}_1},\ 1\ ]  \\
\overrightarrow{\mathbf{b}}^b_{8,q} &= (\mathbf{e}_{41}+\mathbf{a}_{2}) [\ -e^{-i\mathbf{q}\cdot\mathbf{a}_2},\ 0,\ 0,\ 1\ ]  \\
\overrightarrow{\mathbf{b}}^b_{9,q} &= (\mathbf{e}_{32}+\mathbf{a}_{2})[\ 0,\ -e^{-i\mathbf{q}\cdot\mathbf{a}_2},\ 1,\ 0\ ]  \\
\overrightarrow{\mathbf{b}}^b_{10,q} &= (\mathbf{e}_{31}+\mathbf{a}_{1}+\mathbf{a}_{2}) [\ -e^{-i\mathbf{q}\cdot(\mathbf{a}_1+\mathbf{a}_2)},\ 0,\ 1,\ 0\ ].
\end{aligned}
\end{equation}
In a similar way, we can derive the dynamical matrix resulting from the constrain of bond angle. The bond angle deviation can be written as~\cite{mao2013entropic}
\begin{align}
\begin{aligned}
	\Delta \alpha_h&=  \frac{1}{a}\left[\mathbf{e}_{i j} \times\left(\mathbf{u}_{i}-\mathbf{u}_{j}\right)-\mathbf{e}_{k j} \times\left(\mathbf{u}_{k}-\mathbf{u}_{j}\right)\right] \cdot \hat{\mathbf{z}} \\
	& =\frac{1}{a}\left[\mathbf{e}_{i j}^{\perp} \cdot\left(\mathbf{u}_{i}-\mathbf{u}_{j}\right)-\mathbf{e}_{k j}^{\perp} \cdot\left(\mathbf{u}_{k}-\mathbf{u}_{j}\right)\right]
\end{aligned}
\end{align}
where  $a$ is the bond length. $i, j, k$ label the particles forming the bond angle. $\hat{\mathbf{z}}$ is perpendicular to the bond angle plane and $\mathbf{e}_{i j}^{\perp}$ is the unit vector perpendicular to the $\mathbf{e}_{i j}$ in the angle plane. The corresponding dynamic matrix for bond angle bending in Fourier space can be obtained as
\begin{align}
	\mathbf{D}_{\mathbf{q}}^{\text {angle }}=\frac{\kappa}{a^{2}} \sum_{s} \overrightarrow{\mathbf{b}}_{s, \mathbf{q}}^{g} \overrightarrow{\mathbf{b}}_{s,-\mathbf{q}}^{g}
\end{align}
with
\begin{align}
\begin{aligned}
	\overrightarrow{\mathbf{b}}_{s, \mathbf{q}}^{g}= & \mathbf{e}_{l^s_{1} l^s_{2}}^{\perp p^s_{1} p^s_{2}} \otimes\left[ \hat{\mathbf{P}}_{p^s_{1}}-e^{-i \mathbf{q} \cdot\left(\mathbf{r}_{l^s_{2}}-\mathbf{r}_{l^s_{1}}\right)} \hat{\mathbf{P}}_{p^s_{2}}\right] \\
	& -\mathbf{e}_{l^s_{3} l^s_{2}}^{\perp p^s_{3} p^s_{2}} \otimes\left[ \hat{\mathbf{P}}_{p^s_{3}}-e^{-i \mathbf{q} \cdot\left(\mathbf{r}_{l^s_{2}}-\mathbf{r}_{l^s_{3}}\right)} \hat{\mathbf{P}}_{p^s_{2}}\right].
\end{aligned}
\end{align}
The complete dynamic matrix of the crystal lattice is
\begin{align}
	\mathbf{D}_{\mathbf{q}}=\mathbf{D}_{\mathbf{q}}^{\text {bond }}+ \mathbf{D}_{\mathbf{q}}^{\text {angle }}
\end{align}
Its dimensionless counterpart is
\begin{align}
	\widetilde{\mathbf{D}}_{\mathbf{q}}=\beta\sigma^2\left(\mathbf{D}_{\mathbf{q}}^{\text {bond }}+ \mathbf{D}_{\mathbf{q}}^{\text {angle }}\right).
\end{align}
After obtaining the dynamic matrix, we can solve the eigenvalue problem of the dynamic equation
\begin{align}
\partial_t^2 {\mathbf M} \cdot \overrightarrow{\mathbf U}_{l}= - \sum_{l'} {\mathbf D}_{l,l'}  \cdot  \overrightarrow{\mathbf U}_{l'} 
\end{align}
in Fourier space to obtain the phononic spectrum $\omega({\mathbf q} )$, which is essentially the eigenvalue problem of $\widetilde{\mathbf{D}}_{\mathbf{q}}$. Here, ${\mathbf M}$ is the diagonal mass matrix. {In our calculation, we utilize the eigenvalue solver in NumPy library to obtain the phononic spectrums.}

~~

\section*{Associated Content}
{\bf Supporting Information} The Supporting Information is available on the website at DOI: XXXXXX.

{ Section A-C, particle rotational entropy and bond angle rigidity, continuous elastic medium theory, directions of bulk soft modes. Figures S1-S4, the equations of states for different crystal structures at different cutting degree $\alpha$ and temperature, geometric constraint for particles in the \emph{pgg} and \emph{p2-5} structures, the close packed state and the most expanded state without breaking bonds, free energy per particle as a function of $\vartheta$ for \emph{pgg} and \emph{p2-5} structures.}

\ 

{\bf Preprint} Ji-Dong Hu, Ting Wang, Qun-Li Lei, Yu-qiang Ma,	 Transformable super-isostatic crystals self-assembled from segment colloidal rods, \textbf{2023}, \textit{2311.01160}, arXiv, https://doi.org/10.48550/arXiv.2311.01160,  (accessed on 2 Nov 2023)
}

\begin{acknowledgments} 
The authors are grateful to D. Zhou for careful reading of the manuscript and the helpful discussion. This work is supported by the National Key Research and Development Program of China (No. 2022YFA1405000) and the National Natural Science Foundation of China (No. 12275127,  12174184 and  12347102).
\end{acknowledgments}

\bibliographystyle{ACS}
\bibliography{reference}

\clearpage

\begin{figure*}
\centering
		\resizebox{150mm}{!}{\includegraphics[trim=0.0in 0.0in 0.0in 0.0in]{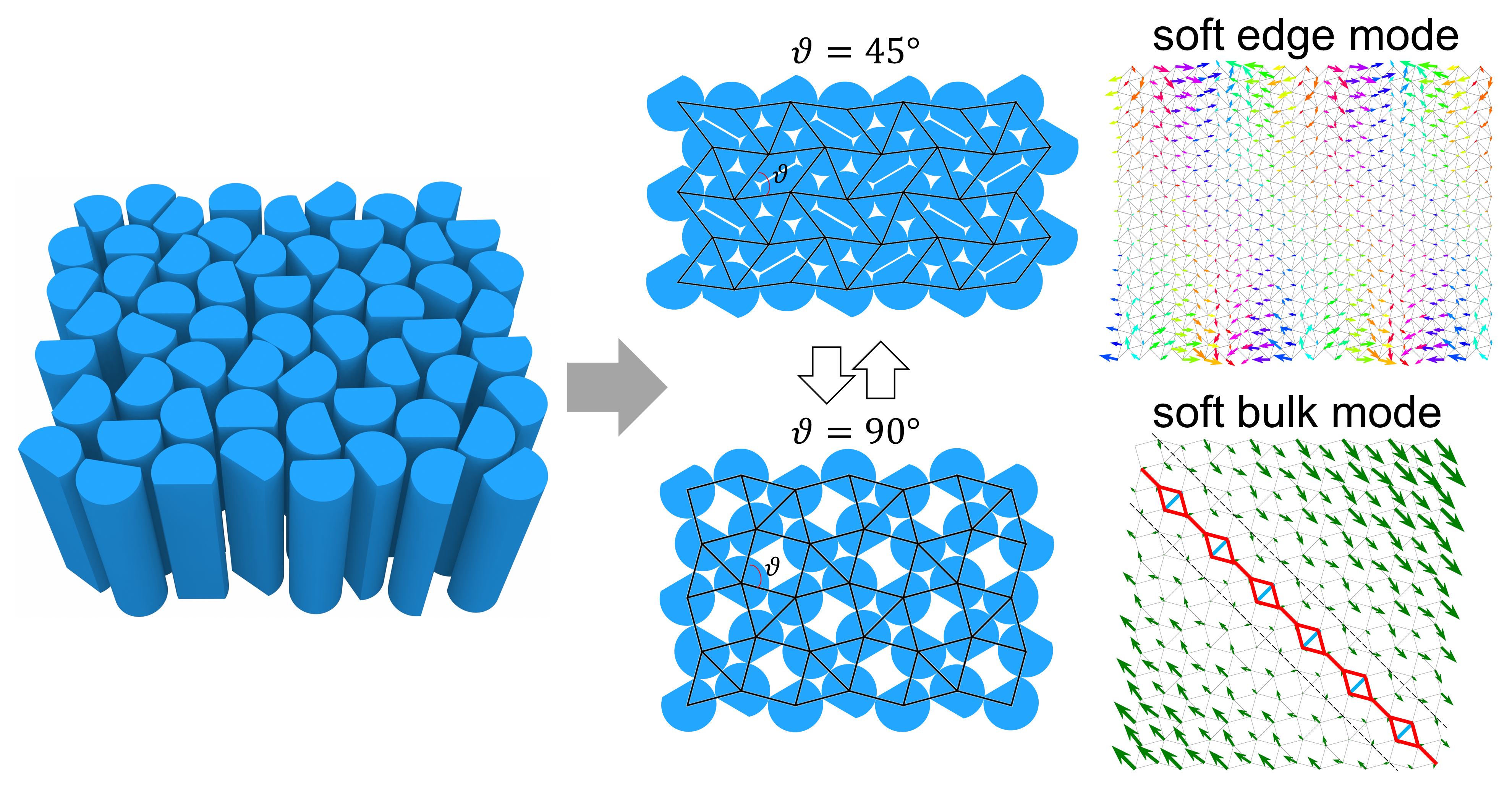} } 
\caption*{  TOC: A nematic monolayer of  segment colloidal rods can self-assemble into a super-isostatic structure capable of transformation between two different mechanical states through a global deformation.   }
\end{figure*}

\end{document}